\documentstyle[11pt,twoside,paspconf,epsf]{article}

\begin{document}

\title{Evolutionary Histories of Dwarf Galaxies in
       the Local Group}

\author{Eva K.\ Grebel\altaffilmark{1}} 
\affil{UCO/Lick Obs., University of California,
    Santa Cruz, CA 95064, USA}

\begin{abstract}
The star formation histories of Local Group (LG) dwarf galaxies and more distant
potential LG members are reviewed.  Problems in defining the spatial
extent of the LG and membership are briefly discussed.
The morphological types found in the LG are presented, and it is suggested 
that we see continuous evolution from low-mass dwarf irregulars (dIrrs)
to dwarf spheroidal galaxies (dSph) in the LG.  

Star formation histories
for LG dwarfs and nearby LG candidates are compiled using population
boxes.  No two dwarfs, irrespective of morphological type, show the same
evolutionary history, and all vary widely in ages of their subpopulations
and in their enrichment history.  The lack of gas in dSphs and certain
dwarf ellipticals (dEs) is puzzling both with respect to their star formation
histories and the expected mass loss from red giants, but a new 
photoionization scenario may reconcile these contradictions.
Old populations, often spatially very
extended, may be a common property of dwarf galaxies, though their fractions
can be very small.  Almost all types of dwarf galaxies studied in detail so far
show spatial variations in ages and abundances such as radial age/metallicity
gradients.  The observed star formation histories 
impose constraints on merger and accretion scenarios.

Properties of the Milky Way dwarf
spheroidals are compared to the M\,31 dSphs and discussed in the framework
of the ram pressure/tidal stripping scenario.  It is demonstrated that the
newly discovered LG dwarfs follow the same relationship for central surface
brightness, mean metallicity, and absolute magnitude as the other LG dwarfs. 
\end{abstract}




\section{Introduction}

Dwarf galaxies are the most frequent type of galaxy in the universe and a 
major constituent of galaxy groups and clusters.  They play an important role
in the formation and build-up of more massive galaxies through mergers and
accretion.  The surviving dwarf galaxies provide a fossil record of the early
formation and evolutionary conditions of their parent clusters.  Galaxies
outside of clusters and groups record the evolution in isolated environments
with a minimum of interactions.  Accurate
knowledge of their star formation and enrichment histories is paramount for
understanding the evolution of clusters and void galaxies
on cosmological time scales.  Combining
evolutionary histories with positional and orbital information helps to
understand the impact of environment and mass on galaxy evolution.

Detailed studies of dwarf galaxy properties are only possible in sufficiently
nearby, resolved galaxies -- the dwarf galaxies of the Local Group (LG), 
whose properties and evolutionary histories will be summarized in this review.  
Other recent reviews of the star formation history of LG dwarfs
include Grebel (1997), Da Costa (1998), Mateo (1998), and van den Bergh (1999). 

For the purpose of this review, old refers to ages $>10$ Gyr,
intermediate-age populations range from 1 to 10 Gyr, and young denotes
populations younger than 1 Gyr. 

\section{Galaxy Content of the Local Group}

We will consider all galaxies with $M_B > -18$ to be dwarf galaxies,
i.e., all LG galaxies other than M\,31, the Milky Way, M\,33,
and the LMC.  However, the known dwarf galaxy census of the LG is
incomplete and its spatial extent poorly defined.  For a comprehensive
picture of LG evolution it is important to understand these limitations. 

\subsection{Morphological types and evolutionary transitions}

We distinguish four basic morphological galaxy types in the LG:
spirals (S), dwarf irregulars
(dIrr), dwarf ellipticals (dE), and dwarf spheroidals (dSph).  
\begin{itemize}
\item The dIrrs are gas-rich, irregularly shaped galaxies with recent or 
ongoing star formation.  Spiral density waves are absent.
\item The dEs are compact, show
very pronounced, dense, bulge-like cores, and may contain gas. 
They contain mainly old and intermediate-age populations and
show in part recent star formation.  Their surface brightness is typically
$\mu_{V,0}  < 21$ mag arcsec$^{-2}$, and $-18 > M_B > -14$.  
\item The dSphs are the least luminous, 
least massive galaxies known and, surprisingly, are almost devoid of gas. 
They do not have a pronounced 
nucleus, show little central concentration and are dominated by old or 
intermediate-age populations.  In contrast to globular clusters
dwarf spheroidals appear to be dark-matter dominated 
(Faber \& Lin 1983, Lin \& Faber 1983) with $6 \la M/L_V \la 100 M_{\odot}$
(Mateo et al.\ 1998).  They are not rotationally supported.
Typically $M \approx
2 \cdot 10^7 M_{\odot}$ (Mateo et al.\ 1998), $\mu_{V,0} > 22$ mag arcsec$^{-2}$
and $M_B > - 14$ (Gallagher \& Wyse 1994). 
\end{itemize}

But are the distinctions between the different morphological types really
so well defined?  dEs and dSphs follow the same scaling relations (e.g.,
Ferguson \& Binggeli 1994) and are often considered the same type of galaxy. 
Low-mass dIrrs may eventually turn into dSphs.  
Transition types (dIrr/dSph) appear to show this evolution in progress.
Like dSphs these galaxies are dominated
by old populations but contain gas and show some recent star formation.
All these intermediate types are at distances $> 250$ kpc from 
the large spirals and are probably less influenced by ram pressure
stripping or tidal stripping.  

\begin{figure}[t]
\plotone{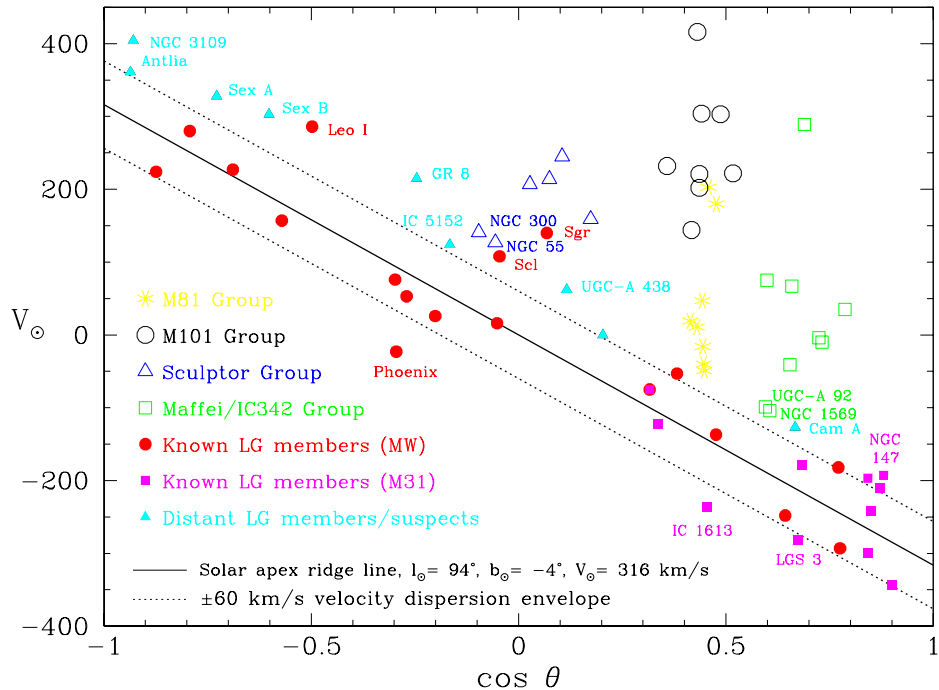}
\vspace{-3.50in}
\caption{Heliocentric velocity, $V_{\odot}$, vs.\ cosine of the angle $\theta$
between solar apex and {\em l,b} of a galaxy.  The solid line represents the
adopted solar motion from Karachentsev \& Makarov (1996).  The dashed lines
indicate the $\pm 60$ km\,s$^{-1}$ velocity dispersion of the Local Group
(Sandage 1986).  Distant Local Group suspects and nearby groups have positive
velocity residuals due to cosmological expansion.  Nearby Local Group dwarfs
with large radial velocities also show positive residuals with respect to
the $\pm 60$ km\,s$^{-1}$ velocity envelope.} \label{fig-1}
\end{figure}

The recent
detection of gas around the Sculptor dSph galaxy (Carignan et al.\ 1998),
the unexpected discovery of a young ($\sim 100$ -- 200 Myr) population in
the dSph Fornax (Stetson et al.\ 1998), and of a $\sim 400$ -- 500 Myr
population in Leo\,I (Gallart et al.\ 1999) changes the traditional
picture of dSphs as gas-devoid galaxies dominated by old or 
intermediate-age populations.
Instead it seems that we see continuous evolution from 
low-mass dIrrs to dSphs, whose details may depend on the mass of the
individual galaxy and its distance from the massive spirals.   
In contrast, blue compact dwarf (BCD) galaxies, which are rotationally
supported systems, are unlikely to evolve into dE/dSph galaxies unless
they can somehow lose their angular momentum (van Zee et al.\ 1998).

In or near the LG additional galaxy types are found:
The dIrr IC\,10 may qualify as a starburst galaxy since its number of 
Wolf-Rayet stars is unusually high for a galaxy of its size, and its WR
star density is twice as high than in any other LG galaxy (Massey \&
Armandroff 1995).  The starburst activity may be triggered by gas
accretion (Wilcots \& Miller 1998).
The dIrr NGC\,3109, a potential LG member depending
on the adopted LG radius, may qualify as a low-surface-brightness spiral.
Very extended spiral structure was detected by Demers et al.\ (1985), 
the surface-brightness is very low (Carignan 1985), and the galaxy's 
rotation curve is dominated by a dark component at nearly all radii 
(Jobin \& Carignan 1990).  While the LG does not contain any 
BCD galaxies, the closest BCD, NGC\,6789, is at a
distance of only 2.1 Mpc (Drozdovsky \& Tikhonov 1999).

\subsection{Dynamical and distance considerations}

The location of 
galaxies within the $\pm 60$ km\,s$^{-1}$ velocity dispersion envelope
in a cos $\theta$, $V_{\odot}$ diagram (cosine of the angle
between solar apex and {\em l,b} of a galaxy versus its heliocentric
velocity: Fig.~1) can be used 
as preliminary membership indicator (e.g., van den Bergh 1994, Grebel 1997).  
This method suggested LG membership for Schmidt's \& Boller's (1992)
LG suspect AM\,1001-270 (Grebel 1997), which was shown 
to be a possible LG member through a distance determination 
by Whiting et al.\ (1997), who named it Antlia after its parent constellation.
For Leo\,A the diagram also predicted membership,
and the revised distance to this dwarf galaxy (Tolstoy et al.\ 1998)
is consistent with that.  On the other hand, nearby dwarf galaxies with 
large radial velocities such as Leo\,I (Zaritsky et al.\ 1989;
see Byrd et al.\ 1994 for an
interesting interpretation), Sculptor, and the currently merging
Sagittarius dwarf spheroidal galaxy (dSph) lie above the LG  
velocity dispersion boundary and would be regarded as non-members if 
distance information were lacking (Fig.~1).
Galaxies with barycentric distances $> 1.2$ Mpc such as 
NGC\,3109, Sex\,A, Sex\,B and GR\,8, which are often considered
LG members also show positive velocity residuals with respect to
the assumed LG locus. In contrast, the positions of IC\,1613 and Phoenix
indicate infall motion.  

While distances give a better indication of LG membership it is not
clear out to which distance galaxies may be considered members of the LG. 
Using the formalism of Yahil et al.\ (1977) and Sandage (1986)
Grebel (1997) derived a radius of 1.8 Mpc for a spherical, barycentric 
zero-velocity surface of the LG.  However, the galaxies of the LG have 
an asymmetric, filamentary distribution (Karachentsev 1996), and 
the gravitational potential of the 
LG is likely to be nonspherical (Whiting, priv.\ comm.).

The simplified assumption of a sphere with 1.8 Mpc radius implies
overlap with other
nearby groups such as the Sculptor group and the Maffei/IC\,342 group.
For instance, at a distance of $\sim 1.65$ Mpc from the barycenter of the
LG NGC\,55, a galaxy usually associated with the Sculptor group,
lies within the zero-velocity surface for the LG (Mateo 1998), 
while dynamical calculations support Sculptor group membership (Whiting 1999).
It is likely that the orbits of more distant LG members 
are affected by tidal forces of nearby groups, while outlying galaxies in 
nearby groups are influenced by the LG.  Dunn \& Laflamme (1993) 
suggest that at redshifts of $z \ge 1$ tidal forces on LG galaxies were dominated
by galaxies outside the LG.  

LG membership can only be established for certain when a galaxy's
orbit is known.  Orbital information based on proper motions, 
however, is available only for six 
nearby Milky Way companions so far.  In this review star formation 
histories of a number of dwarf galaxies out to distances of 1.8 Mpc from the LG 
barycenter will be discussed though galaxies bound to the LG may
fill a smaller volume.
 
\begin{figure}[t]
\plotone{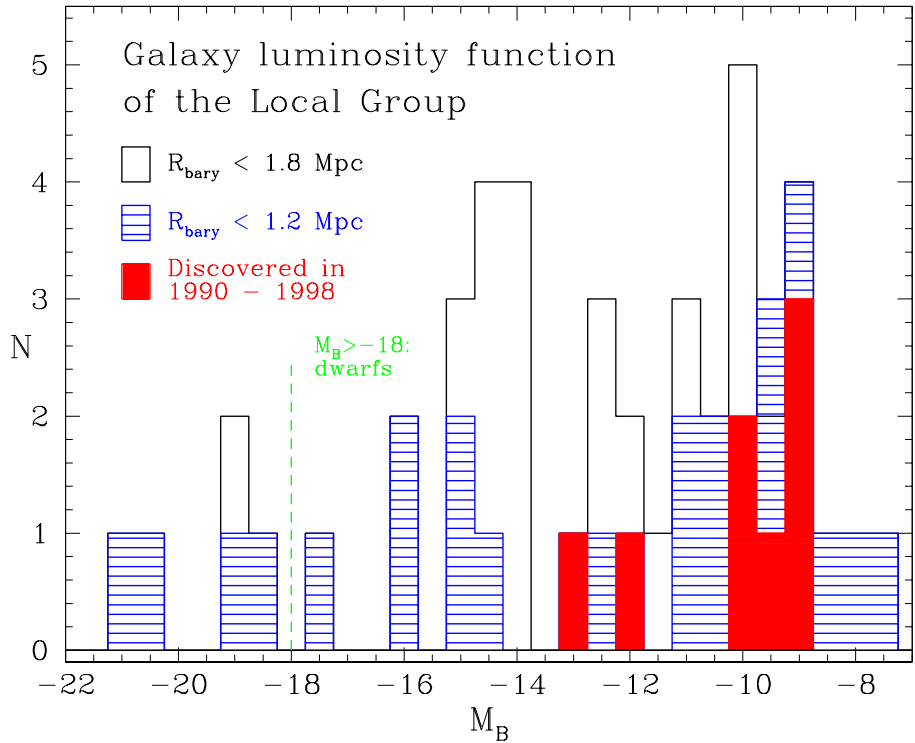}
\vspace{-3.25in}
\caption{$B$-band luminosity function of galaxies within 1.8 Mpc and 1.2 Mpc
from the Local Group barycenter (excluding probable members of nearby groups).
Data on the most recently detected Local Group dwarfs are from
Karachentsev et al.\ (1999) and Grebel \& Guhathakurta (1999).  $M_B$ for
And\,V was estimated following Armandroff et al.\ (1998).  Data for
other galaxies was taken 
from Schmidt \& Boller (1992) and Mateo (1998).  Note how the recent
detections augment the faint end of the luminosity function, though it is still
poorly populated in comparison to other nearby groups and 
clusters.}\label{fig-2}
\end{figure}

\subsection{Incompleteness of the dwarf galaxy census of the Local Group}   

The LG is unique since here the low-mass end of the galaxy luminosity 
function can be studied in great detail.
For example, dSph galaxies, which seem to be
the dominant low-mass galaxy type are undetectable at higher redshifts.
A complete local dwarf census 
might allow us to determine location and properties
of the expected low-mass turnover or termination of the luminosity function.  

The galaxy luminosity functions of nearby galaxy groups and clusters
show a pronounced upturn for $M_B>-14$ mag (Trentham et al.\ 1998a),
while the LG luminosity function (Fig.~2)
is sparsely populated at its faint end.  This may be in part an effect of small
number statistics, but probably also reflects an incomplete local dwarf census.
New faint dwarf galaxies are still being detected in the LG.  In the past
eight years alone eight new potential LG members were discovered (1990: Sextans,
Tucana; 1994: Cam\,A, Sagittarius, 1997: Antlia; 1998: And\,V, And\,VI/Peg\,Dw,
Cas\,Dw).  Five of these are fainter than $M_B=-10$.  Very low overall
luminosity, low surface brightness, vicinity to a bright, superposed foreground
star or a combination of these factors had prevented detection.  Note that the 
only two known dwarf galaxies with $M_B>-8$ mag (Draco \& Ursa Minor) are 
nearby neighbors of the Milky Way.  More still undetected galaxies of similarly
low luminosity may lie at larger distances.   The ongoing survey of 
Karachentseva and Karachentsev (1998) is revealing additional candidates,
which we are following up on with the 2.4-m MDM telescope and the 10-m
Keck telescopes (Guhathakurta et al.\ 1999).  The galaxies studied so far are
all at barycentric distances $> 1.7$ Mpc and unlikely LG members, but they
help to complete the dwarf census in the Local Volume.
The Sloan Digital Sky Survey is expected to further increase the LG galaxy 
census.  Velocity dispersion measurements of yet to be detected
$M_B>-8$ galaxies that are sufficiently distant not to be tidally affected
by the large spirals would show whether dSphs are indeed dark-matter
dominated and whether $M \approx 2 \cdot 10^7 M_{\odot}$ is the lower mass
limit for dSphs.

Does the extreme low-mass end of the luminosity function consist of mere 
H{\sc i} clouds that have not even formed stars yet?  Some of 
the isolated, compact high-velocity
clouds of Braun \& Burton (1999), with masses up to several $10^7$
$M_{\odot}$ and infall motion toward the LG, might be 
progenitors of dwarf galaxies in accordance with the scenario of Blitz et al.\
(1999).  They may also cause, or contribute to,  star formation episodes 
through accretion, as appears to be the case for IC\,10 (see Wilcots \& 
Miller 1998).

\section{Spatial Distribution of Local Group Dwarf Galaxies}

\subsection{Morphological segregation}

The spatial distribution of LG members and LG suspects shows 
morphological segregation (Fig.~3).  An extended, scattered distribution of 
dwarf irregular galaxies (dIrrs) is also observed in other groups (e.g.,
C\^ot\'e et al.\ 1997). The Tucana dSph is currently the only
seemingly isolated dSph known in the LG, while all other dSphs are found
within $< 300$ Mpc of the large spirals.  If this reflects a detection bias,
then deep surveys should uncover additional dSphs at large distances.  On 
the other hand,  a deep survey of the Coma cluster (Trentham 1998b) shows a 
concentration and possibly confinement of the faintest dwarfs toward the
cluster core.  While the steep slope of the faint part of luminosity 
functions in richer clusters agrees with expectations from hierarchical 
cluster formation (Philipps et al.\ 1998), a different scenario has been
proposed for the origin of Milky Way dwarf satellites.  

\subsection{Polar orbital planes?}

The Milky Way's dwarf satellites and a number
of halo globular clusters appear to lie along two (e.g., Majewski 1994)
or more (Lynden-Bell \& Lynden-Bell 1995, Fusi-Pecci et al.\ 1995) 
polar great circles, possibly indicating a common origin as remnants of at 
least two larger parent galaxies since accreted by the Milky Way.  
The available proper motions for LMC, SMC (Kroupa \& Bastian 1997), Draco, 
and Ursa Minor (Scholz \& Irwin 1994) appear consistent with motion along 
the Magellanic stream, while the proper motion of Sculptor (Schweitzer et al.\
1995) neither confirms or rules out motion along the proposed 
Fornax-Leo-Sculptor stream.  Proper motions for the other more distant 
dSphs are not yet available.

\begin{figure}[t]
\plotone{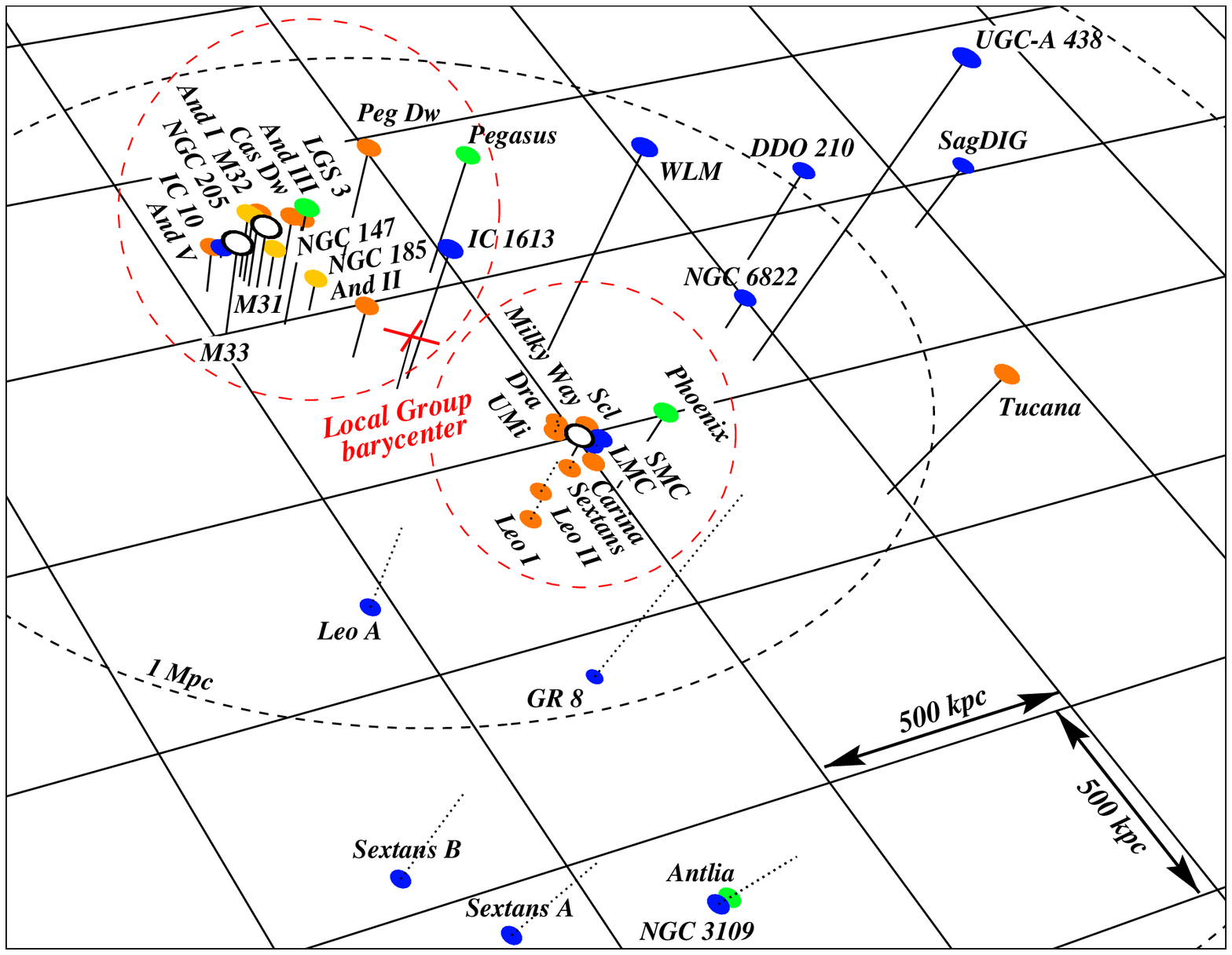}
\vspace{-3.25in}
\caption{A scaled 3-D representation of the Local Group (LG).  
The dashed ellipsoid marks a radius
of 1 Mpc around the LG barycenter (cross).  The underlying grid is parallel 
to the plane of the
Milky Way.  Galaxies above the plane are
indicated by solid lines and below with dotted
lines.  The dashed circles enclose the presumed M\,31/M\,33
and the Milky Way subsystem.  Morphological
segregation is evident:  The dEs and gas-deficient
dSphs (light symbols) are closely concentrated around the large spirals
(open symbols). DSph/dIrr transition types
(e.g., Pegasus, LGS\,3, Phoenix) tend to be somewhat
more distant.  Most dIrrs (dark symbols)
are isolated and located at larger distances.}\label{fig-3}
\end{figure}

Five out of six of M\,31's dSph companions as well as the dIrr/dSph LGS\,3
and dIrr IC\,10 appear to lie within $\pm 15^{\circ}$ of a potential 
Andromeda polar plane when
transformed to a native $l_{M31}$, $b_{M31}$ coordinate system (Grebel
et al.\ 1999a).  For the M\,31 dSphs there are at present not even radial 
velocities available, and the existence of a polar orbital plane is purely
speculative at this point.

\subsection{Constraints on the merger history}
 
In the merger remnant scenario 
one would expect that the epoch at which the disruption occurred
should be marked by similar abundances in the galaxies associated with
each polar plane.  The Magellanic Clouds, Draco, and Ursa Minor can be tidal
fragments of the same parent only if the break-up happened at a very early 
time; else the observed star formation and enrichment histories make a 
common origin unlikely (Olszewski 1998).
Unavane et al.\ (1996) concluded from a study of ages ages and abundances in 
Galactic halo field stars that the major merger epoch must have been 
$\ga 10$ Gyr ago.

Only two of the Milky Way dSphs, Fornax and Sagittarius, contain globular
clusters.
They can contribute both globular clusters of the same age as the
oldest halo globulars (Buonanno et al.\ 1998, Layden \& Sarajedini 1997), and 
clusters up
to 3 Gyr (Fornax GC\#4, Marconi et al.\ 1999) or 7 Gyr younger (Sagittarius's
Terzan\,7, Chaboyer at al.\ 1996). 
The two most metal-poor Fornax globular clusters, GC\#5 ($-2.2$ dex) 
and GC\#1 ($-2.2$ dex), show the second-parameter effect (Smith et al.\
1996, Buonanno et al.\ 1998), but neither Fornax's nor Sagittarius's
globulars resemble the ``young'' halo clusters (Smith et al.\ 1998).

\section{Deriving Star Formation Histories}

\begin{figure}[t]
\plotone{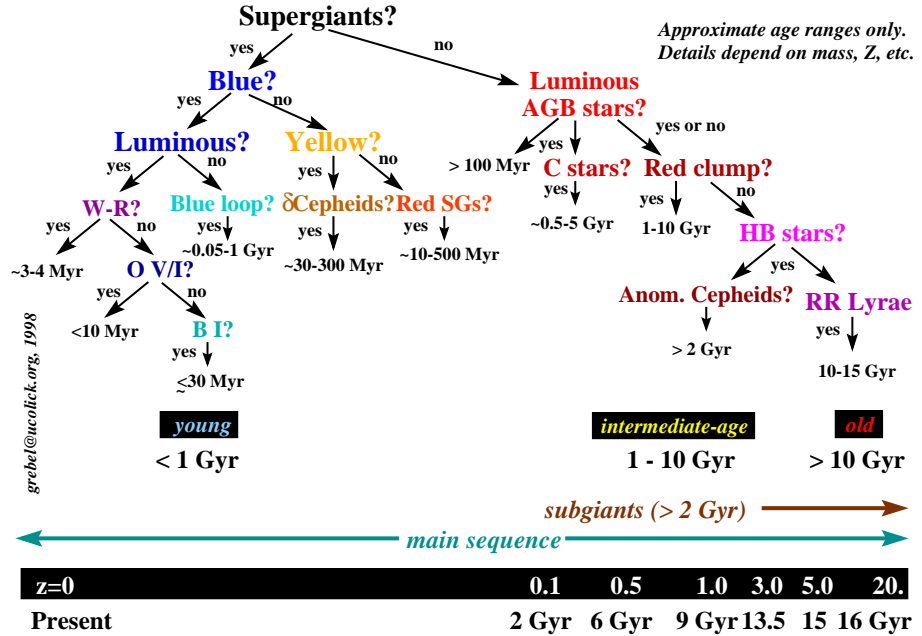}
\vspace{-4.15in}
\caption{Stars as age tracers.  Listed age ranges are rough estimates only.
{\em Note how the resolution decreases drastically 
with increasing age}.  The redshift--age axis 
was derived for H$_0 = 50$ km\,s$^{-1}$\,Mpc$^{-1}$, 
$\Omega = 0.2$.}\label{fig-4}
\end{figure}

Studies of the evolutionary history of a galaxy usually combine 
deep CCD photometry to derive color-magnitude diagrams (CMDs), and 
of stars as age tracers (Fig.~4).  Where available this is supplemented
by spectroscopic abundance information, ISM and kinematic information.

The determination of the star formation history of a galaxy is a
multi-parameter problem.  The interpretation depends crucially on the
knowledge of reddening (both foreground and internal extinction) and distance.
Data may be compromised by crowding and high or variable
internal extinction.  With increasing distance it
becomes more and more difficult to study the properties of the older
populations.  Mixed populations and the age-metallicity-reddening-distance
degeneracy present major challenges and may lead
to ambiguous interpretations.  The ambiguity can be reduced by using
independent estimators such as foreground extinction
from the IRAS/DIRBE reddening maps (Schlegel et al.\ 1998), or by using 
independent spectroscopic abundances. 

Synthetic CMDs, sophisticated modelling techniques and statistical
evaluations permit the reduction of the ambiguities to some extent 
and the extraction 
detailed star formation histories (see Aparicio's 1999 review).
It is important to remember that
these analyses depend on how well evolutionary models reproduce observational
parameters (Fig.~5) as well as on assumptions about IMFs, binary fractions,
and metallicity evolution.  Results obtained with different evolutionary
models may not be directly comparable and tend to differ in ages, age spreads,
and enrichment. 

A way of testing these methods is through the creation of an artificial,
yet realistic
stellar population composed of input data with well-determined (differential)
ages and spectroscopic abundances such as Galactic and Magellanic Cloud 
clusters corrected for distance and reddening.  It would be interesting to
see what star formation histories will be extracted through synthetic CMD 
methods by different groups.  Several groups have expressed
their willingness to participate in such a test, and we hope to carry out
this experiment in the near future. 

Special types of stars can be crucial for uncovering subpopulations
even when high-quality CMDs are available.  For instance, 
the presence of a small intermediate-age population may not be obvious from
the CMD of an old dSph, but the detection of carbon stars
traces this population unambiguously.
RR Lyrae stars or blue horizontal branch stars indicate an old population
(though their lack does not necessarily imply the absence 
of an old population since second-parameter effects can also play a role).

\begin{figure}[t]
\plotone{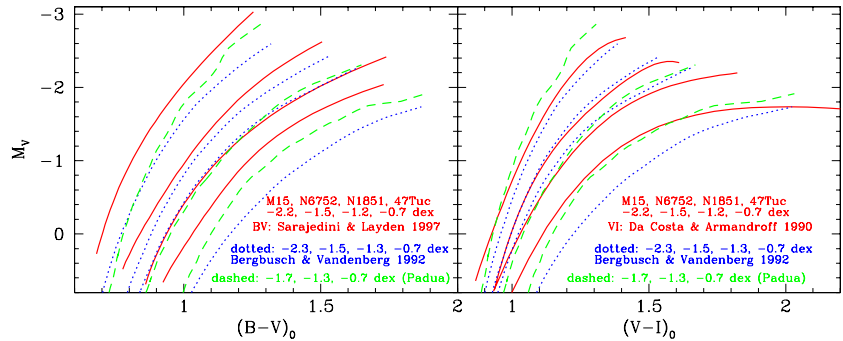}
\vspace{-5.30in}
\caption{Comparison of the mean locations of observed red giant branches 
of Galactic globular clusters (Sarajedini \& Layden 1997, 
Da Costa \& Armandroff 1990) with theoretical isochrones from Bergbusch \&
VandenBerg (1992) and Bertelli et al.\ (1994).  Note the good agreement for
intermediate metallicities ([Fe/H]$ = -1.3$ dex) and the discrepancies for
higher and very low metallicities, which affect the results from
synthetic color-magnitude diagram studies.
}\label{fig-5}
\end{figure}

While photometry is the primary method for stellar population studies,
spectroscopy provides accurate abundances, wind properties, spectral types,
radial velocities, or stellar velocity dispersions.
The new 10-m-class telescopes will make many more
LG galaxies accessible for 
spectroscopic studies and help to uncover
enrichment histories and chemical evolution as well as internal kinematics
and velocity dispersions.  The planned astrometric satellite missions will 
increase and improve astrometric information for LG galaxies and help to 
derive their kinematics and orbits in the LG.  Ultimately this will enable 
us to constrain past interaction events.

Star formation histories must be supplemented by studies of the ISM in and
around galaxies -- gas content and distribution, current star formation
rates, kinematics, abundances (e.g., Young \& Lo 1997ab, Wilcots \& Miller 
1998, Sage et al.\ 1998).
The ongoing HIPASS H{\sc i} multibeam survey (Staveley-Smith 1997) will allow
the mapping of the large-scale distribution of H{\sc i} in the LG and beyond
and to search for extended H{\sc i} clouds associated with dwarf galaxies
(Carignan 1999).  Deep ongoing H$\alpha$ surveys will help to establish
whether photoionized gas (Section 7.3) is associated with dSph galaxies
and whether high-velocity clouds permeate the Local Group (Maloney \&
Bland-Hawthorn 1999).

\section{Star Formation Histories of Local Group Dwarfs}

\begin{figure}[t]
\includegraphics{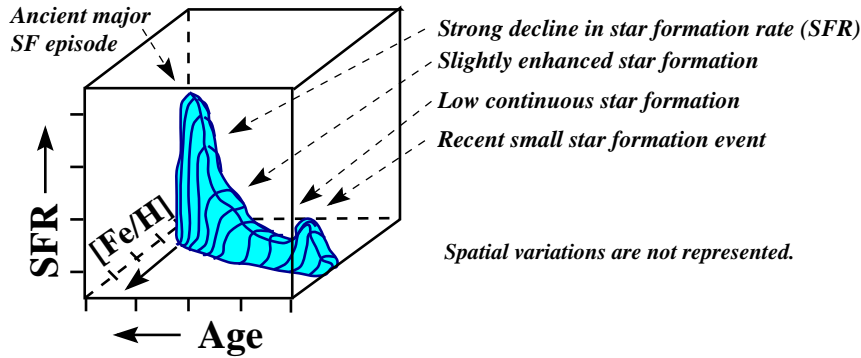}
\vspace{01.90in}
\caption{
A sample population box for a fictitious galaxy.
}\label{fig-6}
\end{figure}

\subsection{Population boxes}

Hodge (1989) introduced population boxes as a
3-D visualization of the
evolutionary history of a galaxy (Fig.~6) as a function 
of time, star formation rate, and metallicity. 
Figures 7 and 8 show a 
compilation of star formation histories of LG galaxies 
based on data 
from sources using heterogeneous observational techniques
and theoretical models  
(see Tab.\ 4 \& 5 in Grebel 1997 for a compilation of properties)

Data on metallicities and enrichment are often uncertain.  A galaxy may have
a metallicity estimate for the old population from the slope of the RGB and
an oxygen abundance determination from an H{\sc ii} region, which can be 
translated into a metallicity value (but see Richer \& McCall 1995 for 
caveats).  In other cases abundance estimates are based solely on synthetic
CMDs and thus depend on the reliability of the input models (see Fig.~5).  
For only a few galaxies are there spectroscopically
determined abundances for a large number of stars.  Substantial metallicity 
spreads 
(e.g., $-3.0 < {\rm [Fe/H]} < -1.5$ dex in Draco, Shetrone et al.\ 1998)
are found in the old populations of some dSphs, 
while other galaxies show very little evidence of metallicity spreads or
enrichment.  A galaxy that experienced hardly any enrichment
despite repeated star formation episodes appears to be Carina (Smecker-Hane
et al.\ 1994).  

Time-dependent star formation rates (SFR) are not yet known
for most galaxies.  Therefore qualitative estimates of relative subpopulation
ratios are used.  It should be remembered that time resolution decreases with 
age and that we have very little information about details of 
star formation episodes longer ago than $\approx 10$ Gyr. 

\begin{figure}[t]
\plotone{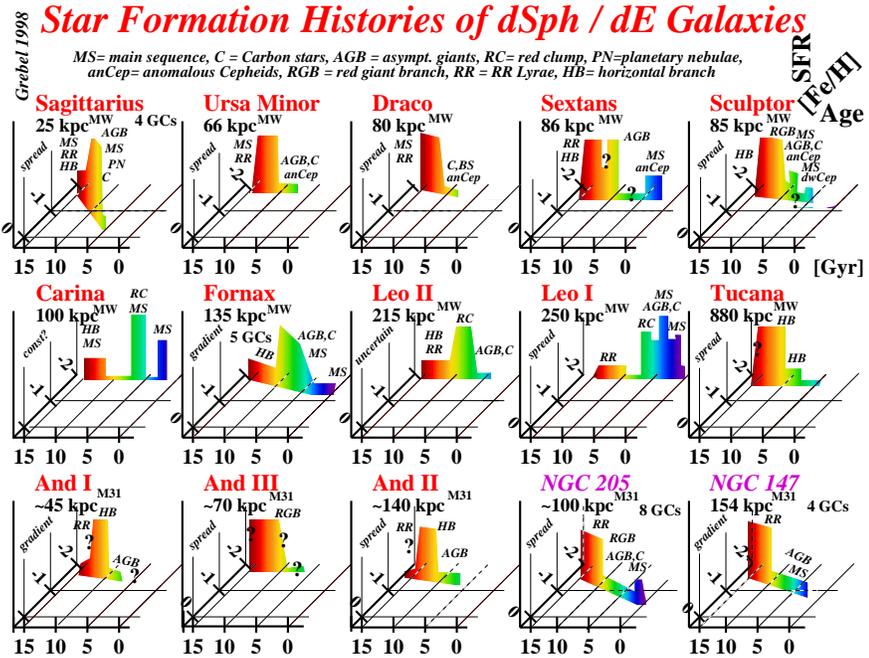}
\vspace{-3.50in}
\caption{
The upper two panels list population boxes for Milky Way dSphs (exception:
Tucana) and their distances from the Milky Way.  The lower panel 
distances refer to M\,31. For references, see  Table 5
in Grebel (1997) and the following more recent studies: Marconi et al.\
1998, Layden \& Sarajedini (1997), Grillmair et al.\ 1998,  Hurley-Keller
et al.\ 1999, Hurley-Keller et al.\ 1998, Mighell 1997,  Stetson et al.\
1998, Stetson \& Grebel 1999, Gallart et al.\ 1999, Da Costa 1998, 
and Armandroff \& Da Costa 1999.  
}\label{fig-7}
\end{figure}


No two dwarf galaxies in the LG have the same star formation
history.  Some show pronounced enrichment, while others remain almost
unchanged despite repeated episodes.  Epochs and time scales of star
formation episodes vary widely.

\subsection{Ram pressure/tidal stripping}

For the 9 dSphs associated with the Milky Way (panels 1 and 
2 in Fig.~7) the dominant populations tend to become younger
with increasing distance from the Milky Way --- a potential consequence
of tidal/ram pressure stripping as suggested by Lin \& Faber (1983)
and van den Bergh (1994).  Interestingly, no such behavior is observed
for the six known M\,31 dSph companions (Grebel \& Guhathakurta 1999),
none of which appears to have a dominant intermediate-age population
even though though they span the same range of galactocentric
distances as the Milky Way dSphs.  Since we have no orbital information,
we cannot exclude that the more distant dSphs are on eccentric orbits
and were stripped of their gas in previous close passages.  Also, M\,31 is more
massive than the Milky Way, and more distant satellites may thus be 
affected by ram pressure stripping.  The strong correlation between
central surface brightness and present-day galactocentric distance found 
for Milky Way dSphs by Bellazzini et al.\ (1996) is much less pronounced 
for the M\,31 dSphs, and their central surface brightness values show a smaller
range in magnitude.  This may reflect the comparatively stronger tidal effects
from M\,31 but is also caused by the lack of dominant 
intermediate-age populations.
Note that the dIrr/dSph LGS\,3 lies at a similar deprojected distance from 
M\,31 and contains a significant intermediate-age population. 
The newly discovered M\,31 companions follow the relationship between
central surface brightness, mean metallicity, and absolute magnitude as the
other LG dSphs, dIrrs, and dEs (Fig.~9; Grebel \& Guhathakurta
1999).

\begin{figure}[t]
\plotfiddle{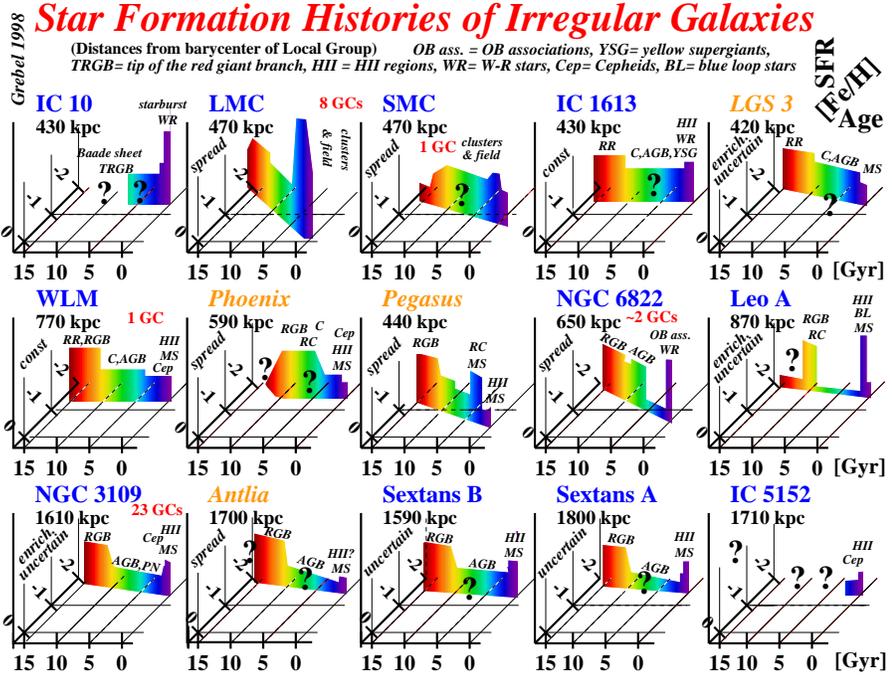}{10.0cm}{0}{70}{70}{-210}{-260}
\caption{
Star formation histories of dIrr and dIrr/dSph (italics) galaxies 
visualized through population boxes.  For references, see Table 4 in
Grebel (1997) and the following more recent studies: Geha et al.\
1998, Grebel et al.\ 1999b, Aparicio et al.\ 1997b, Mould 1997, 
Minniti \& Zijlstra 1997, Mart\'{\i}nez-Delgado et al.\ 1999, Gallagher
et al.\ 1998, Aparicio et al.\ 1997c, Skillman et al.\ 1997, Tolstoy et
al.\ 1998, Whiting et al.\ 1997, Aparicio et al.\ 1997a, Sarajedini et 
al.\ 1997, Dohm-Palmer et al.\ 1997, and van Dyk et al.\ 1998.   
}\label{fig-8}
\end{figure}

\subsection{Old populations -- ubiquitous?}

Old populations have been detected in most galaxies studied in detail so far.
In galaxies dominated by intermediate-age populations (e.g., Leo\,A, Tolstoy
et al.\ 1998) the presence of a small old population cannot be excluded. 
In dIrrs affected by crowding and internal extinction the usually more
extended old populations were detected through deep imaging of the halo 
regions (Minniti \& Zijlstra 1996, Minniti et al.\ 1999).  It seems
that there was a
common epoch of the earliest star formation in at least some dwarf galaxies.   
Main-sequence photometry shows that
the oldest globular clusters in the LMC, Sagittarius, and 
Fornax are as old as the oldest Galactic halo clusters (Olsen et al.\ 1998,
Montegriffo et al.\ 1998, Buonanno et al.\ 1998).  The same has been 
suggested for the oldest globulars in M\,31 and M\,33 (Sarajedini
et al.\ 1998), while the SMC began to form globular clusters $\approx 3$ 
Gyr later (Mighell et al.\ 1998).

\subsection{Spatial variations in metallicity and star formation history}

Most LG dwarfs show position-dependent variations in 
metallicity and star formation history.  For young populations in dIrr 
these variations are easily traced by the spotty distribution of H{\sc ii} 
regions.  {From} an analysis of supergiants and Cepheids Grebel \& Brandner
(1998) found the time scale for continuing star formation in the LMC within
areas of $\la 500$ pc to be of the order of $\la 200$ Myr, similar to time
scales observed in several dIrr galaxies (e.g., Dohm-Palmer et al.\ 1997: 
Sextans\,A; Dohm-Palmer et al.\ 1998: GR\,8). 
In Sex\,A recent star formation progressed from
the center outwards and currently takes place along the inside of an H{\sc i}
ring around the center (van Dyk et al.\ 1998).  Similar central H{\sc i} holes
have also been detected in dIrr galaxies outside the LG (Puche \& 
Westphal 1994).  Depending on the shell propagation velocity, the density
of the ambient medium, and the mass of the parent galaxies the gas may 
eventually be blown out.

Old populations in irregulars and dIrrs (e.g., Sex\,A: Hunter \& Plummer 1996,
van Dyk et al.\ 1998;
WLM: Minniti \& Zijlstra 1996; NGC\,3109: Minniti et al.\ 1999)
have been found to be much more
extended than recent star formation events, which tend to be concentrated
near the center of these galaxies.
Extended old stellar ``halos'' have also been found around around transition
types such as the dSph/dIrr Antlia (Aparicio et al.\ 1997a, Sarajedini et al.\
1997), dSphs like And\,I (Da Costa et al.\ 1996), Carina (Mighell 1997), 
Fornax (Grebel 1997, Stetson et al.\ 1998, Grebel \& Stetson 1999), and 
Sculptor (Hurley-Keller et al.\ 1999), while younger populations are again
confined to the central regions.  

\begin{figure}[t]
\plotfiddle{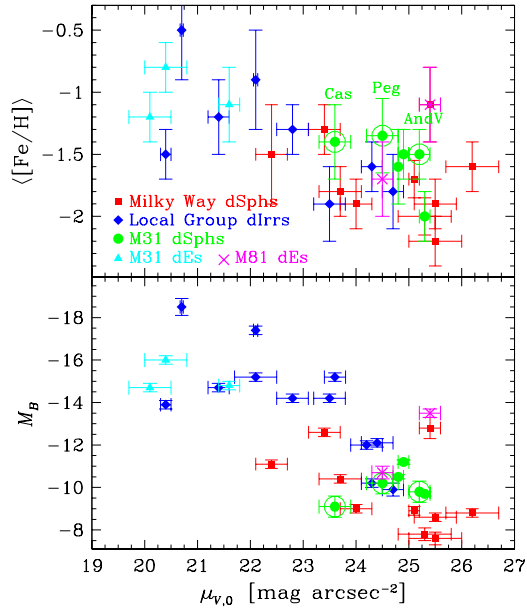}{8.0cm}{0}{50}{50}{0}{-50}
\caption{Mean metallicity, $\rm\langle[Fe/H]\rangle$, plotted
versus central surface brightness in the $V$ band, $\mu_{V,0}$, for nearby
dwarf galaxies (top), and the same for absolute $B$-band magnitude, $M_B$, 
versus $\mu_{V,0}$ (bottom).  
The data are from Grebel (1997), Armandroff et~al.\ (1998),
Grebel \& Guhathakurta (1999), Mateo (1998), and Karachentsev \& Karachentseva 
(1998).  The points representing the newly 
discovered M\,31 companions are circled.  Note that they follow the 
mean $\rm\langle[Fe/H]\rangle$-vs-$\mu_{V,0}$-vs-$M_B$  relation within the
errors. 
} \label{fig-9}
\end{figure}

\section{Modes of Star Formation}

In low-mass galaxies the initial star formation episode
may cause most of the gas to be expelled
while more massive galaxies are able to retain their
gas much longer or even continue to form stars over a Hubble time 
(Hunter 1997).  

\subsection{Continuous star formation}

Slow, continuous star formation interrupted by short, inactive periods
(``gasping star formation'') was found in a number of the more massive
LG dIrrs (NGC\,3109: Greggio et al.\ 1993; Sex\,A: Aparicio et al.\ 1987;
Sex\,B: Tosi et al.\ 1991, NGC\,6822: Marconi et al.\ 1995, Gallart et al.\
1996).  Despite winds from massive stars and supernovae these galaxies are
sufficiently massive to retain much of their gas and to undergo gradual 
enrichment (De Young \& Heckman 1994) through 
feedback from newly formed stars and gas recycling. 

\subsection{Interactions, accretion, and triggered star formation}

Accretion plays an important role in providing star-forming material.
Star formation with metal-poor accreted material is a possible explanation
for the distinct episodes observed in the Carina dSph (Smecker-Hane et 
al.\ 1994, Hurley-Keller et al. 1998), the long pauses in between, and
the lack of enrichment.

Young \& Lo (1997b) and 
Welch et al.\ (1998) found the gas in the dE NGC\,205 to be rotating in 
contrast to the stars in this galaxy.  Interaction with a gas cloud in the
disk of M\,31 or accretion of an intergalactic H{\sc i} cloud are likely
explanations and may be responsible for the recent episode of star formation
in NGC\,205.  

The dIrr IC\,10 shows counter-rotating H{\sc i} gas in its outer parts and
is embedded in a very extended H{\sc i} cloud (Wilcots \& Miller 1998),
both likely signatures of recent and still ongoing accretion.  The current 
starburst was probably triggered by this accretion event.  Apart from
extended tidal tails (e.g., Putman et al.\ 1998) and mutual disruption
both Magellanic
Clouds show enhanced cluster formation 100--200 Myr ago (Grebel et al.\ 1999b), 
coincident with with the closest encounters between the Clouds, and with
the Milky Way (Gardiner \& Noguchi 1996).  



\section{Gas in/around Dwarf Ellipticals and Dwarf Spheroidals}

\subsection{Neutral gas}

NGC\,185, a galaxy with ongoing star formation, is the only one of the four 
dE companion of M\,31 whose observed 
H{\sc i} content is consistent with expectations from stellar mass loss
(Sage et al.\ 1998). 
NGC\,147 with predominantly old and intermediate-age populations 
(Han et al.\ 1997)
contains less than 2\% of the expected gas (Sage et al.\ 1998) and represents
the most extreme case of gas deficiency among the dEs.  Supernova explosions,
interactions with, and passages through, the disk of M\,31 (Sofue 1994)
may have stripped the M\,31 companions of part of their gas.
As discussed by Sage et al.\ (1998),
neither differences in mass nor evolutionary history can explain the 
observed differences in the gas content of the four M\,31 dEs. 
If NGC\,147 and NGC\,185 are a jointly formed, bound binary pair of   
comparable mass (van 
den Bergh 1998), the differing gas and dust content is even more puzzling.  

The H{\sc i} content of LG dSphs has upper limits of $10^2$ --
$10^5 M_{\odot}$ (Knapp et al.\ 1978, Thuan \& Martin
1979, Mould et al.\ 1990, Koribalski et al.\ 1994).  While a few supernova
events may suffice to eject most of the gas of a low-mass dSph
(Mac Low \& Ferrara 1998), these
upper limits are even below the expected return from red giant mass loss.  
The lack of gas is particularly surprising in dSphs with repeated, 
well-separated episodes of intermediate-age star formation
such as in Carina (Smecker-Hane et al.\ 1994, Hurley-Keller et al.\ 1998)
or very recent star formation such as in Fornax (Stetson et
al.\ 1998) and Leo\,I (Gallart et al.\ 1999).   

Gas with $> 3 \cdot 10^4 M_{\odot}$
moving with the systemic velocity of Sculptor has been detected
around this galaxy (Carignan et al.\ 1998; see also Knapp et al.\ 1978).
The amount of gas is consistent with 10\% of the estimated
mass loss from RGB stars during the past 8--10 Gyr and may either have
been ejected in an earlier star formation episode, stripped by Galactic
tides (indeed the highest concentrations are found along Sculptor's
direction of motion), or could be of external origin.  The HIPASS survey
is currently being used to search other dwarf galaxies for similar 
surrounding H{\sc i} structures.  Gas clouds rather than shells 
were detected near two other dwarf galaxies (Tucana: Oosterloo et al.\ 1996,
Sagittarius: Bland-Hawthorn et al.\  1998), but it is unclear whether these
clouds are associated with the dwarfs. 

\subsection{Heated/photoionized gas}

Type Ia supernovae may heat and disperse gas, which remains confined by the 
dark matter halo of a dwarf galaxy (Burkert \& Ruiz-Lapuente 1997).  After
several Gyr the gas will have cooled down sufficiently to initiate a second 
epoch of star formation.  The new generation should have higher metallicity
after having been enriched by the Type Ia supernovae.  While such time gaps
are observed in Carina, the younger populations do not seem to have been 
enriched.

Lin \& Murray (1999) suggest a scenario  
in which newly formed massive stars
and supernovae quench further star formation and ionize the residual gas,
which expands into the outer regions of the dwarf galaxy.  It is retained even
if it is outside the tidal radius (Oh et al.\ 1995), and continues to follow
the orbit of the dwarf. The low density extends the cooling time beyond 1 Gyr
and the gas remains ionized
through the extragalactic UV radiation field and radiation from the nearest 
spiral.  The gas converges with the host galaxy at apogalacticon, where the
exposure to the Galactic UV flux is minimized.  After one or several 
apogalactic passages the gas may become sufficiently compressed to cool,
recombine, and eventually form stars again.  The Lin \& Murray 
scenario accounts for the non-detection of dSphs in H{\sc i} and 
for long dormant periods without necessarily implying enrichment.  Deep
H$\alpha$ Fabry-P\'erot surveys are currently on their way to test the 
prediction of photoionized gas around dSphs.   
 

\section{Summary}

The size of the LG and consequently its membership are poorly defined.  
The outer regions of nearby groups overlap with the outskirts of the LG.  
Knowledge of orbits is required to understand interactions with nearby groups
and within the LG itself, and to reliably constrain membership.

The LG dwarf galaxy census is incomplete, and the faint end of the LG
galaxy luminosity function is deficient in comparison to other nearby
groups.  Is the morphological segregation
in the LG caused to some extent by a detection bias?  Are there more 
distant, isolated dSphs like Tucana to be discovered?  

In 1998 the discovery of
three new dwarf galaxies was announced, all of which turned out to be
distant dSph companion candidates of M\,31 (Armandroff et al.\ 1998,
Karachentsev \& Karachentseva 1998,
Tikhonov \& Karachentsev 1999, Grebel \& Guhathakurta 1999). 
The new dSphs follow the same relationship for central surface
brightness, mean metallicity, and absolute magnitude as the other LG dwarfs.
In contrast to Milky Way dSphs at comparable distances from the parent spiral
these dSphs do not show indications of dominant intermediate-age populations
as one might have expected from the ram pressure/tidal stripping scenario
(Lin \& Faber 1983, van den Bergh 1994), nor do they show the strong 
correlation between central surface brightness and galactocentric radius
found for Milky Way dSphs (Bellazzini et al.\ 1996).
Sensitive large-scale surveys
may help to uncover more distant, low-surface-brightness LG members. 

The compilation of LG dwarf star formation histories shows that no two
galaxies have the same history, not even within the same morphological
type.  All LG dwarfs vary widely in star formation histories, metallicity
evolution, ages, times and time scales of star formation, and number and
distribution of their subpopulations.  Many dwarf galaxies show spatial
variations in ages and metallicities.  Typically star formation lasted
longest in the central regions of the galaxies, while in the outer parts
often extended old populations are found.  Most galaxies appear to share
a common epoch of star formation $>10$ Gyr ago, but the fractions of these
old populations vary and are difficult to age-date.  Orbital information 
would be
most valuable in order to search for correlations between star formation
episodes and close encounters.  

There may be no firm distinction between different morphological types of
dwarfs in the LG.  Instead the recent detection of young populations in
Fornax (Stetson et al.\ 1998) and Leo\,I (Gallart et al.\ 1999) as well as
the observation of various transition phases argues  
in favor of a continuous evolution from low-mass dIrrs to dSphs.   

Two modes of star formation are observed in the Local Group:  Continuous
(at constant star formation rate or declining) and episodic star formation.
The lack of gas in dEs and dSphs remains puzzling, but possibly associated
gas is being detected in the surroundings of some of the dwarfs. 
According to Lin's \& Murray's (1999) scenario gas is present, but photoionized
and therefore not detected in H{\sc i} studies.  
Tidal compression at apogalacticon may trigger
efficient cooling and onset of star formation, and coincides with the time
scales observed in episodic star formation.  Observational tests of this
scenario are currently being carried out by searching for H$\alpha$ emission.


\acknowledgments

I thank Puragra Guhathakurta for a fruitful and enjoyable
collaboration,  Igor Karachentsev, Valentina Karachentseva,
and Doug Lin for communications prior to publication,
and Taft Armandroff, Claude Carignan, Serge Demers, George 
Jacoby, and Alan Whiting for valuable discussions.
Many thanks to Wolfgang Brandner for his help with IDL.  
I am indebted to Ian Thompson for a critical reading of the text and for his 
patience during the process of the writing of this paper.  
I gratefully acknowledge support by Dennis Zaritsky through NASA LTSA
grant NAG-5-3501 and thank him for making it possible
for me to attend the Symposium.

Extensive use of NASA's Astrophysics Data System 
Abstract Service (ADS) has been made in writing this review.

\bigskip

\centerline{\rule{3cm}{0.2mm}}

\bigskip

\noindent
{\em This review tries to summarize the state of knowledge as of 
early Dec 1998.}

\end{document}